\documentclass[english]{article}
\usepackage[T1]{fontenc}
\usepackage[latin9]{inputenc}
\usepackage{babel}
\usepackage{amsmath}
\usepackage{amssymb}
\usepackage{graphicx}
\usepackage[unicode=true,pdfusetitle,
 bookmarks=true,bookmarksnumbered=false,bookmarksopen=false,
 breaklinks=false,pdfborder={0 0 1},backref=section,colorlinks=false]
 {hyperref}

\makeatletter

\providecommand{\tabularnewline}{\\}

\newenvironment{lyxcode}
{\par\begin{list}{}{
\setlength{\rightmargin}{\leftmargin}
\setlength{\listparindent}{0pt}
\raggedright
\setlength{\itemsep}{0pt}
\setlength{\parsep}{0pt}
\normalfont\ttfamily}%
 \item[]}
{\end{list}}

\usepackage{authblk}
\usepackage{babel}
\usepackage{lineno}

\makeatother

\begin{document}
\title{Fast, accurate solutions for curvilinear earthquake faults and anelastic strain}

\author[1]{Walter Landry\thanks{Corresponding author.
\textit{Email Address:} wlandry@caltech.edu (W. Landry).}\thanks{Present Address: Walter Burke Institute for Theoretical Physics, Caltech, Pasadena, CA 91125, USA}}

\author[2]{Sylvain Barbot\thanks{sbarbot@ntu.edu.sg}\thanks{W. Landry: Methodology, Software, Visualization, and Writing of the paper. S. Barbot:  Methodology, Software, and Writing of this paper.}}

\affil[1]{IPAC, Caltech, Pasadena, CA 91125, USA}

\affil[2]{Earth Observatory of Singapore, 50 Nanyang Avenue, Nanyang Technological University, 639798, Singapore}

\maketitle
\begin{abstract}
Imaging the anelastic deformation within the crust and lithosphere
using surface geophysical data remains a significant challenge in
part due to the wide range of physical processes operating at different
depths and to various levels of localization that they embody. Models
of  Earth's elastic properties from seismological imaging combined
with  geodetic modeling may form the basis of comprehensive rheological
models of Earth's interior. However, representing the structural complexity
of faults and shear zones in numerical models of deformation  still
constitutes a major difficulty. Here, we present numerical techniques
for high-precision models of deformation and stress around both curvilinear
faults and volumes undergoing anelastic (irreversible) strain in a
heterogenous elastic half-space. To that end, we enhance the software
Gamra \cite{landry+barbot16} to model triangular and rectangular
fault patches and tetrahedral and cuboidal strain volumes. This affords
a means of rapid and accurate calculations of elasto-static Green's
functions for localized (e.g., faulting) and distributed (e.g., viscoelastic)
deformation in Earth's crust and lithosphere. We demonstrate the correctness
of the method with analytic tests, and we illustrate its practical
performance by solving for coseismic and postseismic deformation following
the 2015 Mw 7.8 Gorkha, Nepal earthquake to extremely high precision. 
\end{abstract}

\section{Introduction}

The deformation of the Earth's lithosphere at time scales relevant
to the earthquake cycle is accommodated by slip on faults and distributed
strain in finite regions of anelastic deformation. Faults exhibit
complex dynamics governed by nonlinear friction laws \cite{dieterich79a,ruina83}
that involve a wide range of thermally activated chemical and physical
processes \cite{ditoro+04,nakatani01}. In addition, fault networks
can exhibit great geometrical intricacies, and single fault segments
are often characterized by morphological gradients \cite{hubbard+15,hubbard+16}.
The combination of constitutive and geometrical complexities constitutes
a fundamental challenge for earthquake prediction \cite{oglesby08}. 

Crustal and lithosphere dynamics is also characterized by distributed
anelastic deformation. For example, in the lower crust or the mantle
asthenosphere, it is responsible for loading faults and accommodating
transient strains \cite{masuti+16}. The kinematics of crustal deformation
can be inferred from geodetic or seismic data by discretization of
fault surfaces and linear inversions of deformation data during the
interseismic, coseismic, and postseismic phases of the earthquake
cycle \cite{mcguire+03,murray+segall05,bartlow+11,barbot+13}. The
development of these techniques in the last few decades has led to
an explosion of knowledge on fault behavior \cite{rogers+03,bakun+05,bletery+14,wallace+17,araki+17}.
These methods have recently been extended to incorporate the distributed
deformation of large domains of the lithosphere \cite{lambert+barbot16,barbot+17,moore+17,qqiu+18a},
such that it is now possible to build models of Earth's deformation
that represent fault slip and distributed strain consistently using
elasto-static Green's functions. 

Numerical simulations of fault dynamics using Green's function and
the integral method have produced predictive scenarios of seismicity.
Forward models of crustal dynamics are important to reveal the frictional
or rheological properties of the Earth, and to make useful predictions
about the long-term behavior of the mechanical system. A large body
of work focuses on the interaction between localized and distributed
deformation \cite{kato02,benzion+93,lambert+barbot16,devries+16,devries+17}.
These studies rely on analytic or numerical methods that treat fault
slip and viscoelastic flow in fundamentally different ways. The introduction
of Green's functions for distributed deformation allow us to simplify
the treatment of rheological heterogeneities, nonlinear anelastic
behavior, and strong mechanical feedbacks (for example, between shear
heating and rheology). Analytic solutions for rectangular and triangular
dislocations \cite{chinnery61,chinnery63,savage+hastie66,sato+matsuura74,okada85,okada92,nikkhoo+15,gimbutas+12,meade07a}
and cuboidal strain volumes \cite{lambert+barbot16,barbot+17} allow
us to build kinematic and dynamic models of lithosphere deformation
on curved surfaces with distributed, off-fault anelastic strain. Despite
being of fundamental importance, all these solutions share an important
caveat as they do not represent the lateral variations of elastic
properties of the ambient rocks.

The goal of this paper is to describe a reliable numerical method
for constructing displacement and stress Green's functions for curvilinear
faults  and strain volumes. The method builds upon a three-dimensional,
adaptive, multi-grid elasticity solver for embedded dislocations described
 in a recent publication \cite{landry+barbot16}. This manuscript
presents functional improvements of the method to model localized
and distributed deformation in an arbitrary heterogeneous elastic
half space. We adopt the approach used in the geodesy community  and
consider fault surfaces discretized in triangular or rectangular elements.
 Similarly, we consider volumes discretized in tetrahedral or cuboidal
volumes. We provide a short treatment of the modeling approach in
Section \ref{sec:Methods}. We demonstrate the relevance of the technique
by modeling the coseismic and postseismic deformation on the 2015
Mw 7.8 Gorkha, Nepal earthquake in Section \ref{sec:Gorkha}. The
algorithms described in this paper are implemented in Gamra, a freely
available tool for realistic earthquake modeling (see Section \ref{sec:Computer-Code-Availability}).

\section{Methods\label{sec:Methods}}

\subsection{Elasto-statics with embedded dislocations}

We assume that Earth's deformation at static equilibrium is governed
 by the equations of linear elasticity
\begin{equation}
\sigma_{ji,j}+f_{i}=0\,,\label{eq:elasto-statics}
\end{equation}
where $\sigma_{ij}$ is the Cauchy stress and $f_{i}$ is a forcing
term. This assumption  is thought to be valid for seismo-tectonic
activity with infinitesimal strain.  The stress components $\sigma_{ji}$
are defined using Hooke's law in terms of the displacement components
$v_{i}$, Lame's first parameter $\lambda$, and the shear modulus
$\mu$ as 
\begin{equation}
\sigma_{ji}\left(\vec{v}\right)\equiv\mu(v_{i,j}+v_{j,i})+\delta_{ij}\lambda v_{k,k\,}.\label{eq:Stress}
\end{equation}
We use Einstein summation notation, where each index $i$, $j$, $k$
is understood to stand for $x$, $y$, and $z$ in turn, repeated
indices are summed, and commas (,) denote derivatives.  Eqs. \ref{eq:elasto-statics}
and \ref{eq:Stress} govern the static displacement field induced
by fault slip \cite{steketee58a,steketee58b}, or the quasi-static
velocity field driven by viscoelastic flow or other anelastic deformation
processes \cite{barbot+09a,barbot+fialko10b}. In the case of  quasi-static
deformation, the forcing term is per unit time, and $v_{i}$ is the
velocity field. Because a large breadth of physical processes can
be captured within the same functional form, solving equations \ref{eq:elasto-statics}
and \ref{eq:Stress} with realistic material properties can find a
broad range of applications.

\begin{figure}
\begin{centering}
\includegraphics[width=0.3\paperwidth]{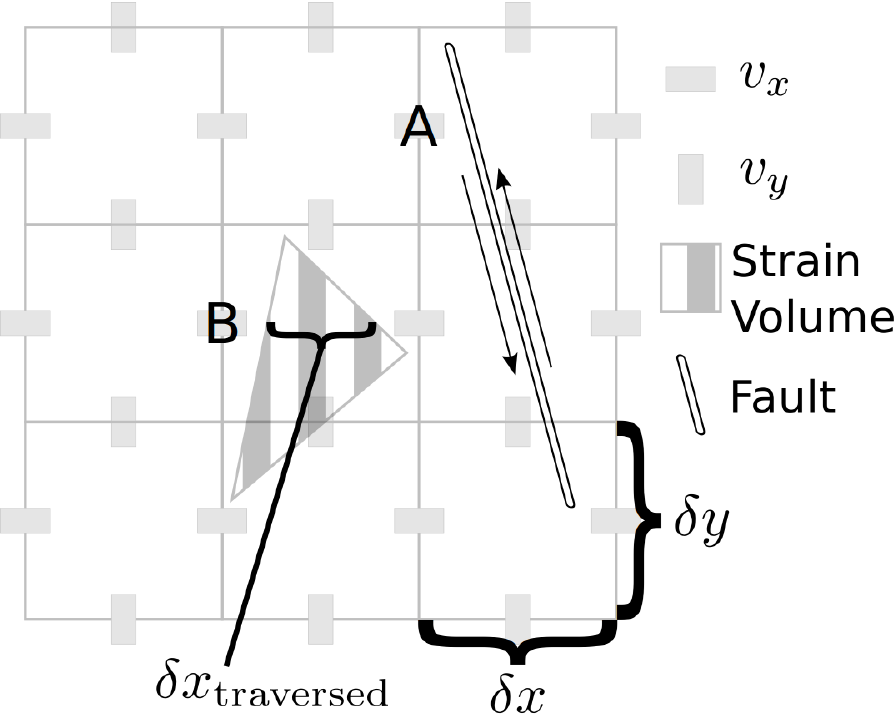} 
\par\end{centering}
\caption{\label{fig:Fault-corrections}Corrections for dislocations and strain
volumes on a staggered grid. This figure is correct for both a 2D
grid and a slice of a 3D grid. In 3D, the dislocations are triangles
or rectangles, but in both cases their intersection with the slice
is a line segment. The 3D strain volumes can be either tetrahedra
or cuboids, so their intersections with the slice are either triangles
or rectangles. The stencil for $\left(\mu v_{x,x}\right)_{,x}$ at
point $A$ gives rise to a correction proportional to the slip $s_{x}$.
At point $B$, the correction is proportional to $\epsilon_{xx}\delta x_{\text{traversed}}$.}
\end{figure}

The method we use to solve these equations is largely described in
the previous paper \cite{landry+barbot16}. To summarize, we use a
parallel multigrid solver on a staggered, adapted finite difference
grid as in Figure \ref{fig:Fault-corrections}. For dislocations,
we add carefully constructed correction terms that depend on the mesh
size. For example, when Eq. \ref{eq:elasto-statics} is expanded out,
it includes the term $\left(\mu v_{x,x}\right)_{,x}$. Expressing
this derivative at point $A=\left(A_{x},A_{y}\right)$ in Figure \ref{fig:Fault-corrections}
using standard finite differences gives

\begin{multline}
\left.\left(\mu v_{x,x}\right)_{,x}\right|_{A_{x},A_{y}}=\left[\left.\mu\right|_{A_{x}+\delta x/2,A_{y}}\left(\left.v_{x}\right|_{A_{x}+\delta x,A_{y}}-\left.v_{x}\right|_{A_{x},A_{y}}\right)\right.\\
\left.-\left.\mu\right|_{A_{x}-\delta x/2,A_{y}}\left(\left.v_{x}\right|_{A_{x},A_{y}}-\left.v_{x}\right|_{A_{x}-\delta x,A_{y}}\right)\right]/\delta x^{2},\label{eq:finite_difference}
\end{multline}
where $\delta x$ is the width of each finite difference cell and
the notation $\left.\mu\right|_{x,y}$ denotes the value of $\mu$
at the coordinate $\left(x,y\right)$. We model dislocations as step
functions $s_{i}$ in displacement $v_{i}$. In Figure \ref{fig:Fault-corrections},
a dislocation with a discontinuity $s_{x}$ in $v_{x}$ passes between
the points $\left(A_{x}+\delta x,A_{y}\right)$ and $\left(A_{x},A_{y}\right)$.
As the resolution $\delta x$ goes to zero, the dominant term in the
finite difference derivative from Eq. \ref{eq:finite_difference}
at point $A$ is

\begin{equation}
\left.\mu\right|_{A_{x}+\delta x/2,A_{y}}s_{x}/\delta x^{2}\,,\label{eq:divergent_derivative}
\end{equation}
which diverges. This divergence is a numerical artifact. The derivatives
are well behaved on both sides of the dislocation, it is only the
numerical approximation that is divergent.

To fix this issue, we consider dislocations as a boundary between
different domains. To use finite difference expressions like Eq. \ref{eq:finite_difference},
we have to transport all of the quantities to the same domain as where
we want to evaluate the expression. For the example in Eq. \ref{eq:finite_difference},
the expression is being evaluated at $\left(A_{x},A_{y}\right)$.
So the terms $\left.v_{x}\right|_{A_{x},A_{y}}$ and $\left.v_{x}\right|_{A_{x}-\delta x,A_{y}}$
are already on the same side. The term $\left.v_{x}\right|_{A_{x}+\delta x,A_{y}}$
has to be transported by subtracting the slip $s_{x}$. This results
in a correction term which is equal and opposite to the divergent
term in Eq. \ref{eq:divergent_derivative}. This is a constant term
that does not depend on the solution, so we can fold this correction
into a modified forcing term
\begin{equation}
\left.\delta f_{x}\right|_{A_{x},A_{y}}=-s_{x}\left.\mu\right|_{A_{x}+\delta x/2,A_{y}}/\delta x^{2}.\label{eq:forcing_correction}
\end{equation}
For interpolation stencils for multigrid, we can use the same logic
to transport all of the quantities to the same domain. The end result
is that the interpolation stencils acquire some extra constant terms.

\subsection{Dislocations on triangular patches}

Our previous work \cite{landry+barbot16} was applied to dislocations
defined on rectangular patches. To extend the method to curvilinear
faults, we assume that a fault surface can be represented by a triangular
mesh where slip is piecewise uniform. This representation is commonly
utilized in geodetic imaging \cite{mcguire+03,loveless+meade10,bartlow+11,qqiu+16}
and earthquake cycle modeling \cite{matsuzawa+10,shibazaki+11,qqiu+16}.
Higher-order discretization methods, such as the non-uniform rational
basis spline, are not yet widely adopted in the field of crustal dynamics. 

\begin{figure}
\begin{centering}
\includegraphics[width=0.4\paperwidth]{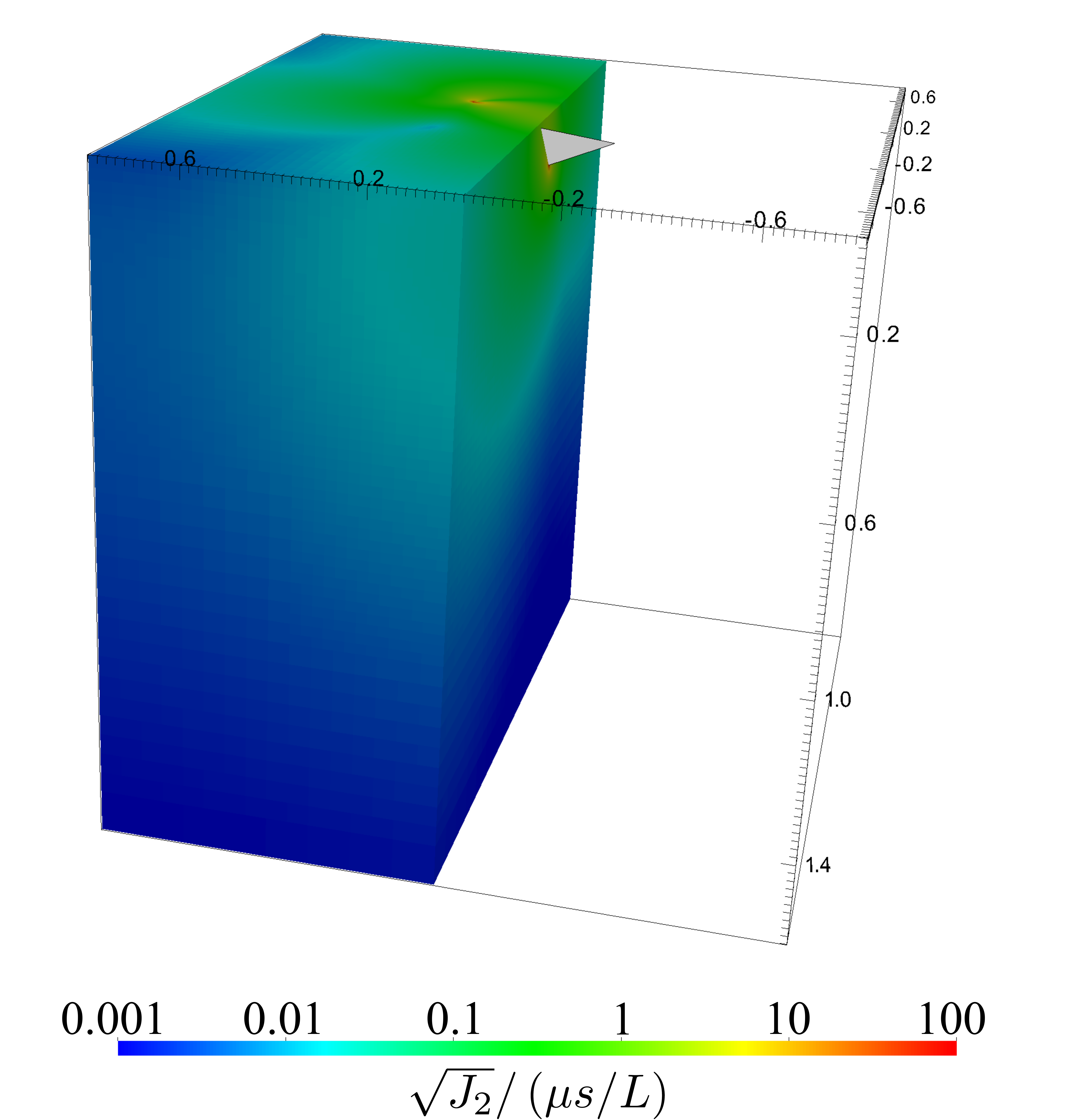}
\par\end{centering}
\caption{\label{fig:triangle-J2}A cutout of the second invariant of the scaled
deviatoric stress $J_{2}=\left(\sigma_{ij}\sigma_{ji}-\sigma_{ii}\sigma_{jj}/3\right)/2$
of a computed solution for a single triangular fault in 3D. The equivalent
resolution of the finest level is $2048\times2048\times2048$. The
fault, indicated in grey, is inclined about 25 degrees from vertical,
has slip $s=10$, and has dimensions $L=0.50$, $W=0.25$. The moduli
are constant ($\mu=\lambda=1$). We set the boundary conditions (normal
Dirichlet and shear stress) from the analytic solution in \cite{nikkhoo+15}.}
\end{figure}

\begin{figure}
\begin{centering}
\includegraphics[width=0.4\paperwidth]{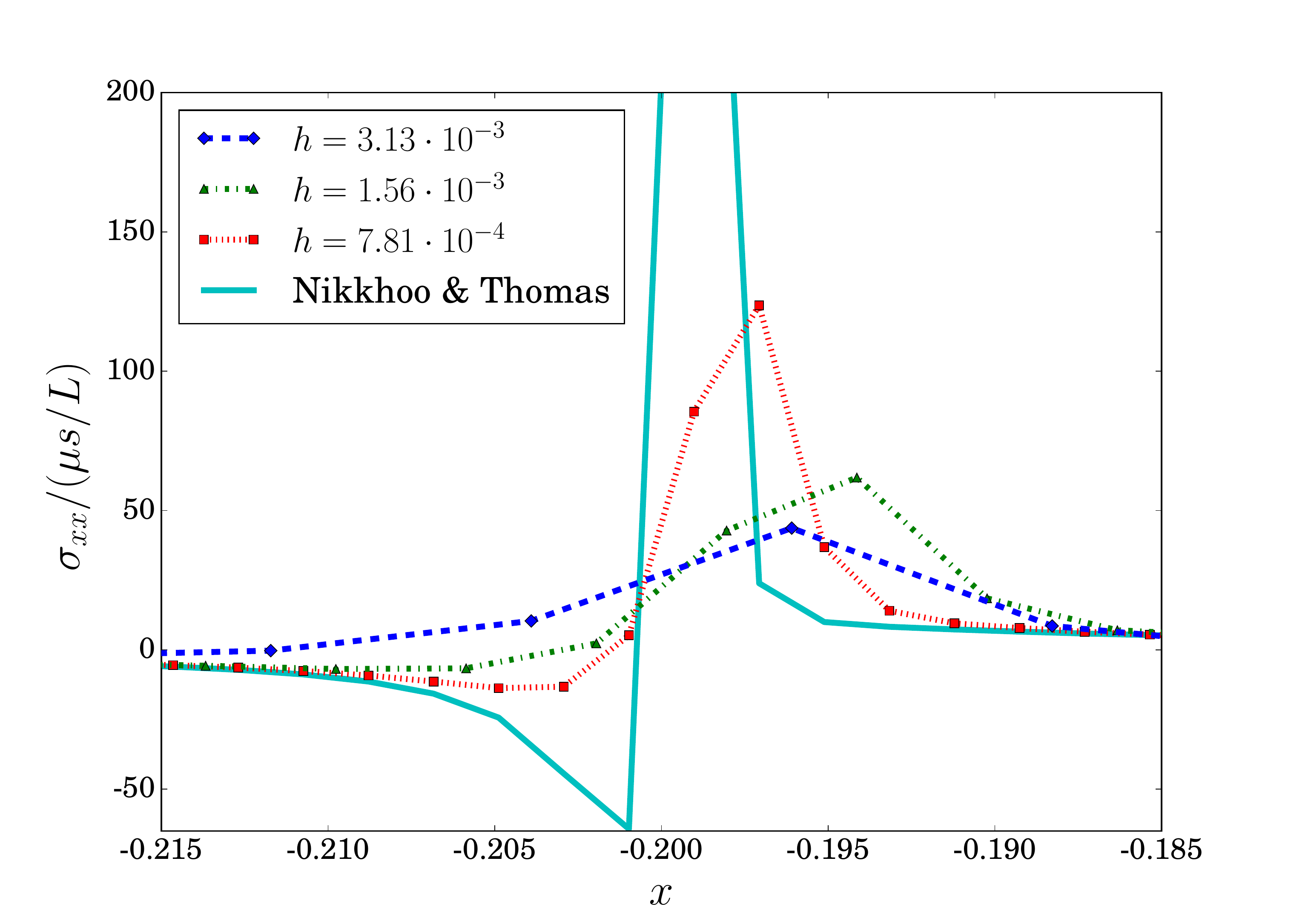}
\par\end{centering}
\caption{\label{fig:triangle_stress}Numerical and analytic solutions for the
scaled stress $\sigma_{xx}/\left(\mu s/L\right)$ due to a single
triangular fault for various resolutions. We set the characteristic
length $L=0.5$ as the longest side of the triangle. The points are
plotted along the line $y=-0.2+h/2$, $z=h/2$, passing near the singularity
at $\left(-0.2001,-0.2001,0\right)$. The points are offset by $h/2$
because of the staggered mesh. The analytic solution from \cite{nikkhoo+15}
is plotted along the same line as the finest resolution.}
\end{figure}

This representation allows us to apply most of the same methods used
for rectangular patches. The correction terms of the finite difference
scheme for Eq. \ref{eq:forcing_correction} are calculated using intersections
between a triangle and stencil elements as in Figure \ref{fig:Fault-corrections}.
 Figure \ref{fig:triangle-J2} shows a cross section of a solution
for a single triangular fault. Figure \ref{fig:triangle_stress} shows
that, even for a line crossing near a singular corner, the computed
stress is well behaved and converges to the analytic solution for
a homogeneous half-space \cite{nikkhoo+15}. This result gives us
confidence that our numerical approach can capture displacement and
stress in  the near field of triangular fault elements with high accuracy. 

\subsection{Strain volumes}

We extend the method to strain volumes to represent distributed anelastic
deformation, such as viscoelasticity, poroelasticity, and thermoelasticity.
 Under the infinitesimal strain assumption, the plastic deformation
of the  lower crust or the asthenosphere can be represented by a linear
combination of finite-volume elements. This allows the simulation
of complex forward models of deformation with nonlinear constitutive
properties using the integral method \cite{lambert+barbot16} or directly
imaging plastic flow in large domains of Earth's interior using surface
geodetic data \cite{moore+17,qqiu+18a}. We are interested in modeling
the displacement and stress in the half-space surrounding these elements
while incorporating a realistic distribution of elastic moduli. To
allow realistic volume meshes,  we consider tetrahedra as the basic
element, but we also include cuboids  for convenience. We assume that
these finite volumes undergo anelastic  deformation with a piecewise
uniform applied strain $\epsilon_{ij}$. A realistic distributed strain
can be modeled by adding up many of these individual elements.

The implementation of internal dislocation and anelastic strain differ
because  for the latter there are no longer step functions in displacement,
but rather gradients. Consider Figure \ref{fig:Fault-corrections}.
The difference between $v_{i}$ at $\left(B_{x}+\delta x,B_{y}\right)$
and at $\left(B_{x},B_{y}\right)$ depends on how much of the strain
volume is traversed by the stencil. In this case, the difference for
$v_{x}$ is $\epsilon_{xx}\delta x_{\text{traversed}}$. Using this,
we compute a correction for strain volumes in a manner similar to
that used to compute Eq. \ref{eq:forcing_correction}, but modified
as follows 
\begin{equation}
\left.\delta f_{x}\right|_{A_{x},A_{y}}=-\epsilon_{xx}\delta x_{\text{traversed}}\left.\mu\right|_{A_{x}+\frac{\delta x}{2},A_{y}}/\delta x^{2}.\label{eq:strain_slip_correction}
\end{equation}
Comparing this with Eq. \ref{eq:forcing_correction}, the only difference
is replacing $s_{x}$ with $\epsilon_{xx}\delta x_{\text{traversed}}$.
This means that after computing $\epsilon_{ij}\delta x_{\text{traversed}}^{j}$
for all of the strain volumes, we can reuse all of the machinery for
adaptive multigrid solutions that we previously employed for dislocations.

\begin{figure}
\begin{centering}
\includegraphics[width=0.4\paperwidth]{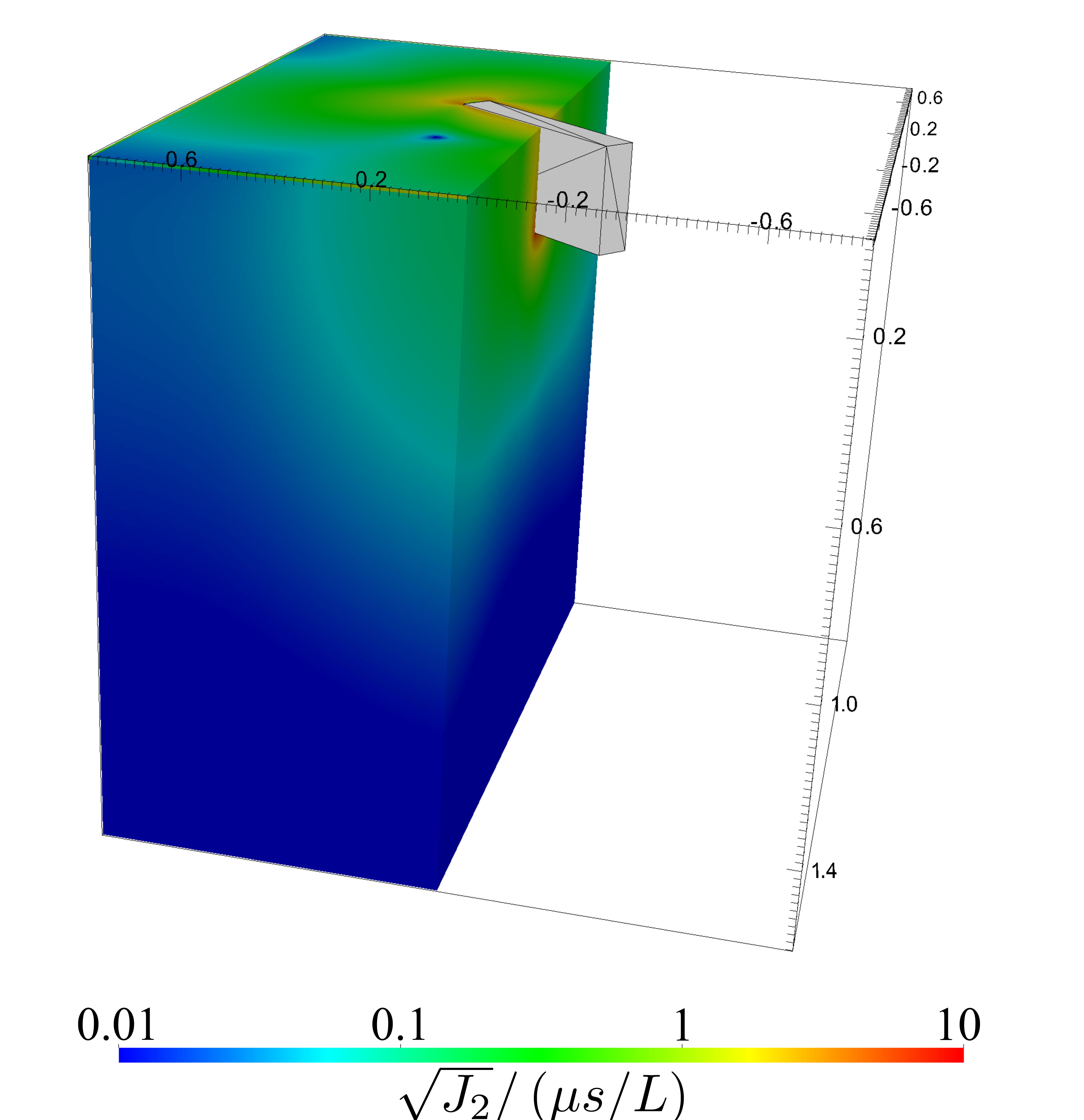}
\par\end{centering}
\caption{\label{fig:strain-J2}A cutout of the invariant of the scaled deviatoric
stress $J_{2}=\left(\sigma_{ij}\sigma_{ji}-\sigma_{ii}\sigma_{jj}/3\right)/2$
of a computed solution for a single cuboidal strain volume. The equivalent
resolution of the finest level is $256\times256\times256$. The strain
volume, indicated in grey, is made up of 6 tetrahedra and has dimensions
$L=0.50$, $W=0.25$, $H=0.08$. Each tetrahedra has shear strain
components $\epsilon_{LH}=\epsilon_{HL}=1600$, so the total slip
across the narrow part of the block is $s=\epsilon_{LH}\times H=1600\times0.08=128$.
The moduli are constant ($\mu=1.6$, $\lambda=1.5$). We set the boundary
conditions (normal Dirichlet and shear stress) from the analytic solution
in \cite{lambert+barbot16,barbot+17}.}
\end{figure}

To test our implementation, there is no analytic solution for the
deformation of tetrahedral volumes in a half space. However, there
is a solution for a cuboid aligned with the surface \cite{lambert+barbot16,barbot+17}.
So we construct a cuboid out of 6 tetrahedra. Figure \ref{fig:strain-J2}
shows the computed solution for this arrangement, and Table \ref{tab:vx_error_strain}
details the convergence of the $L_{\infty}$ error in $v_{x}$. We
also implement cuboids directly, and the solutions are identical to
the solutions of cuboids made up of tetrahedra. 

\begin{table}[h]
\begin{centering}
\begin{tabular}{|l|l|}
\hline 
$h$ & $L_{\infty}\left(\delta v_{x}\right)$\tabularnewline
\hline 
\hline 
0.1 & 26.54\tabularnewline
\hline 
0.05 & 3.191\tabularnewline
\hline 
0.025 & 1.545\tabularnewline
\hline 
0.0125 & 1.222\tabularnewline
\hline 
0.00625 & 0.4033\tabularnewline
\hline 
\end{tabular}
\par\end{centering}
\caption{\label{tab:vx_error_strain}The $L_{\infty}$ norm of the error of
$v_{x}$ for a computed solution of a cuboid strain volume. The strain
volume has shear strain components $\epsilon_{LH}=\epsilon_{HL}=1600$
and dimensions $L=0.50$, $W=0.25$, $H=0.08$. The moduli are constant
($\mu=\lambda=1$). We set the boundary conditions (normal Dirichlet
and shear stress) from the analytic solution in \cite{lambert+barbot16,barbot+17}.
The displacement is regular, so the convergence is uneven but monotonic.}
\end{table}

\begin{figure}
\begin{centering}
\includegraphics[width=0.4\paperwidth]{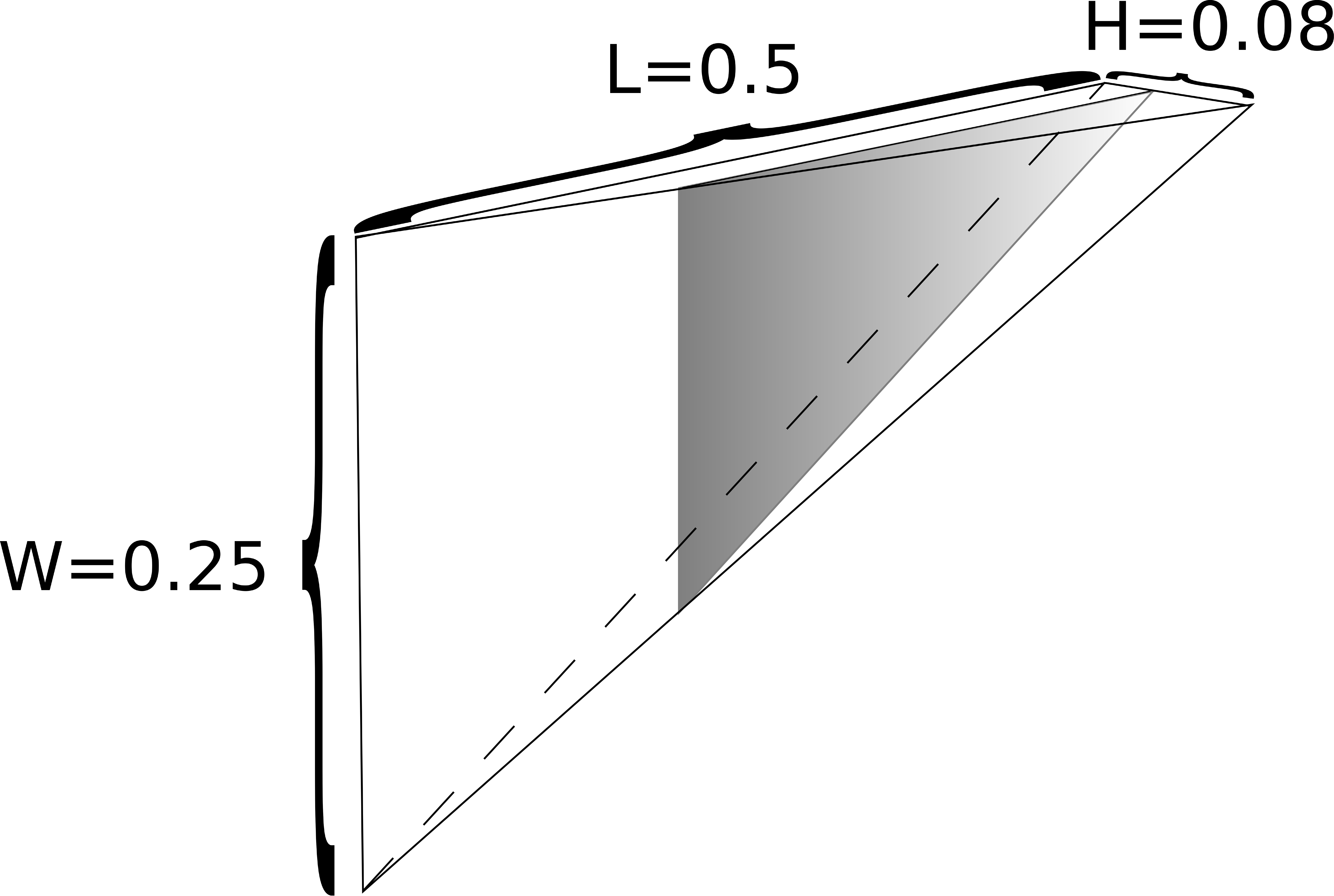}
\par\end{centering}
\caption{\label{fig:tetra_mesh}Dimensions of the single tetrahedra used for
testing variable moduli in strain volumes. To verify the implementation,
we also approximate the strain volume with a large number of triangular
faults distributed along the $H$ axis. The shaded triangle is a representative
single fault. So the triangular faults start out as big as the side
of the tetrahedra and then converge to a single point. }
\end{figure}

\begin{figure}
\begin{centering}
\includegraphics[width=0.4\paperwidth]{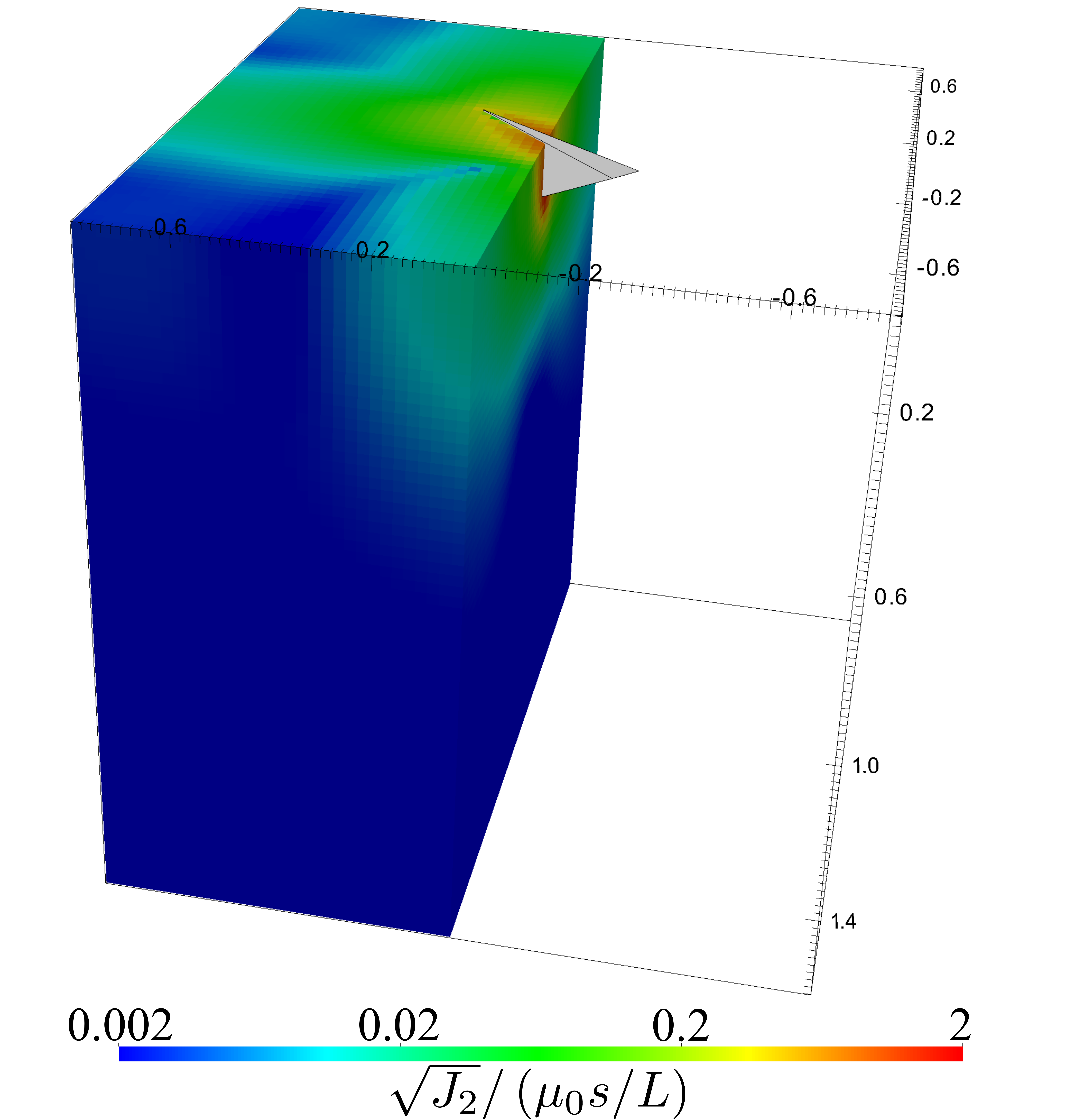}
\par\end{centering}
\caption{\label{fig:strain3D_tetra}A cutout of the invariant of the scaled
deviatoric stress $J_{2}=\left(\sigma_{ij}\sigma_{ji}-\sigma_{ii}\sigma_{jj}/3\right)/2$
of a computed solution for a single tetrahedral strain volume. The
equivalent resolution of the finest level is $64\times64\times64$.
The tetrahedra is the same as one of the tetrahedra making up the
block in Figure \ref{fig:strain-J2}, and has a single non-zero shear
strain component $\epsilon_{LH}=1600$. The moduli are not uniform,
but set to $\mu=\mu_{0}+3x+5y$, $\lambda=\lambda_{0}-2x+7y$, where
$\mu_{0}=10$ and $\lambda_{0}=10$. The boundary conditions are zero
normal displacement and zero stress.}
\end{figure}

To test more complicated volumes with variable moduli, we create a
model using one of the tetrahedra from Figure \ref{fig:strain-J2}.
Figure \ref{fig:tetra_mesh} shows the dimensions of the tetrahedra,
and Figure \ref{fig:strain3D_tetra} shows a solution for the stress.
We do not have an analytic solution for this setup, but we can approximate
a tetrahedral strain volume with a number of small triangular faults
as in Figure \ref{fig:tetra_mesh}. This approach works very well
at coarse resolution. At higher resolution, the mesh can see the spaces
between the faults, so the details start to differ more significantly.
Table \ref{tab:vx_fault_diff} demonstrates that the two methods converge
as the number of faults increases. This gives us some confidence in
the correctness of our implementation of strain volumes  with heterogeneous
elasticity.

\begin{table}
\begin{centering}
\begin{tabular}{|l|l|}
\hline 
$N$ & $L_{\infty}\left(\delta v_{x}\right)$\tabularnewline
\hline 
\hline 
128 & 0.4129\tabularnewline
\hline 
256 & 0.1645\tabularnewline
\hline 
512 & 0.0750\tabularnewline
\hline 
1024 & 0.0520\tabularnewline
\hline 
\end{tabular}
\par\end{centering}
\caption{\label{tab:vx_fault_diff}The $L_{\infty}$ norm of the difference
in $v_{x}$ between solutions that use a tetrahedral strain volume
directly and one approximated with $N$ small triangular faults. The
setup is the same as in Figure \ref{fig:strain3D_tetra}. The triangular
faults are evenly distributed along the width ($H=0.08)$ as shown
in Figure \ref{fig:tetra_mesh}. The slip on each triangular fault
is $\epsilon_{LH}\times H/N$. }
\end{table}

\section{The 2015 Mw 7.8 Gorkha, Nepal earthquake\label{sec:Gorkha}}

As an illustration of the relevance and performance of the proposed
modeling approach, we construct realistic models for the coseismic
deformation and initial post-seismic relaxation of the 2015 Mw 7.8
Gorkha, Nepal earthquake. This earthquake \cite{avouac+15,galetzka+15,elliott+16}
took place on the Main Himalayan Front, a megathrust that separates
the Indian and Eurasian plates \cite{tapponnier+86}. The continental
collision is responsible for the highest mountain ranges on Earth
and has the potential for very destructive earthquakes \cite{bilham95,sapkota+13,bollinger+14,bollinger+16}.
The 2015 Gorkha earthquake only broke a fraction of the megathrust,
prompting questions about the dominant role of fault morphology \cite{qqiu+16,hubbard+16}
or friction \cite{michel+17} in partial ruptures. The main shock
was followed by a detectable transient deformation \cite{wang+fialko16,gualandi+16}
that was compatible with accelerated fault slip in the down-dip extension
of the rupture and viscoelastic flow in the lower crust of Southern
Tibet \cite{zhao+17}. The absence of coseismic and postseismic slip
up-dip of the rupture, i.e., south of Katmandu, has raised serious
concerns about the remaining seismic hazard in the region \cite{bilham15}.
We simulate the static deformation that accompanies coseismic fault
slip on the megathrust and the velocity induced by viscoelastic flow
in the lower crust of the down-going plate. In both cases, the solution
encompasses a box $600\times600\times300$ 
km. The boundary conditions on the side and bottom are free slip:
zero shear stress and zero normal displacement. The boundary conditions
on the top are free surface: zero shear and normal stress. The adapted
mesh ranges from a resolution of 9,300 m down to 146 m. This is equivalent
to a fixed resolution of $4096\times4096\times2048$. A uniform mesh
would require about $3\times10^{10}$ elements, but the adapted mesh
only requires $4-6\times10^{7}$ elements. The models took between
16 and 26 hours to solve on a Dell R720 with 16 physical cores (Intel
Xeon CPU ES-2670).

\subsection{Coseismic slip on a curved fault}

\begin{figure}
\begin{centering}
\includegraphics[width=0.6\paperwidth]{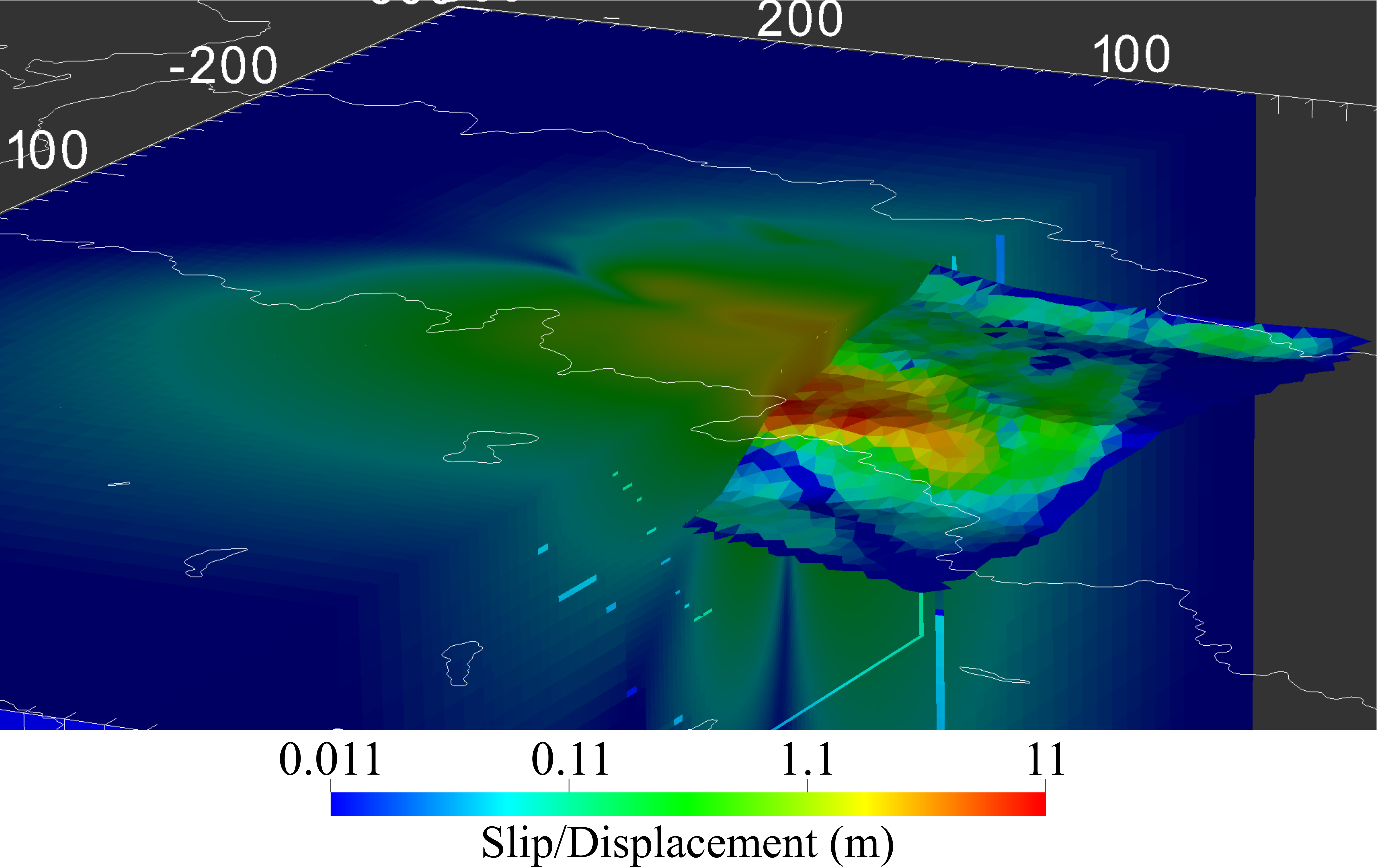}
\par\end{centering}
\caption{\label{fig:Gorkha_slip}A cutout of the applied slip and induced displacement
for the 2015 Mw 7.8 Gorkha, Nepal earthquake. The scale of the model
is marked in km, and the political boundaries of Nepal are superimposed
for reference.}
\end{figure}

We use the slip model from \cite{qqiu+16}, which consists of 2,841
triangular patches with individual slip amplitudes and directions.
The slip model incorporates a realistic megathrust geometry constrained
by structural and geological observations \cite{hubbard+16} and explains
a comprehensive geodetic dataset made of spaceborne radar observations
and GPS measurements \cite{qqiu+16}. The elastic moduli are set with
PREM \cite{dziewonski+anderson81}. Figure \ref{fig:Gorkha_slip}
shows the slip model and computed displacements. Because of the low-angle
décollement, the deformation is confined above the hanging wall and
above the rupture. Despite the level of detail packed into the model,
the simulation required no manual meshing. The regions of highly variable
displacement strain are re-meshed adaptively \cite{landry+barbot16}.
To push the limits of the method, we set the refinement critera to
refine when the error in the displacement is greater than 0.1 cm.
This is about 0.01\% of the maximum displacement, and about 10 times
smaller than the error from having the boundaries only 300 km away. 

\subsection{Distributed postseismic deformation}

The stress change induced by the main shock can accelerate viscoelastic
flow in ductile regions. Remarkably, the governing partial differential
equations for the instantaneous velocity field due to viscoelastic
flow are the same as for elasto-static deformation when equivalent
body forces (per unit time) are used to represent the irreversible
deformation \cite{barbot+09a,barbot+fialko10b}. This equivalence
can be exploited to map out the distribution of strain rate in Earth's
interior using the surface velocity field to reveal, for example,
the spatial distribution of effective viscosity \cite{moore+17,qqiu+18a}. 

We illustrate the capability of computing the instantaneous velocity
field due to viscoelastic deformation in the lower crust of the down-going
plate. To generate the strain rate volumes, we assume that the lower
crust dives 35 km down a ramp centered on the Main Himalayan Front
\cite{cattin+avouac00}. We create a mesh of the lower crust using
$15\times15\times5$ km cuboids. Using the slip model of \cite{qqiu+16},
we use the analytic solution \cite{nikkhoo+15} to compute the change
in stress at the center of each cuboid. We then set the strain rate
of each cuboid assuming an effective viscosity of $10^{20}$ Pa$\cdot$s.

\begin{figure}
\begin{centering}
\includegraphics[width=0.6\paperwidth]{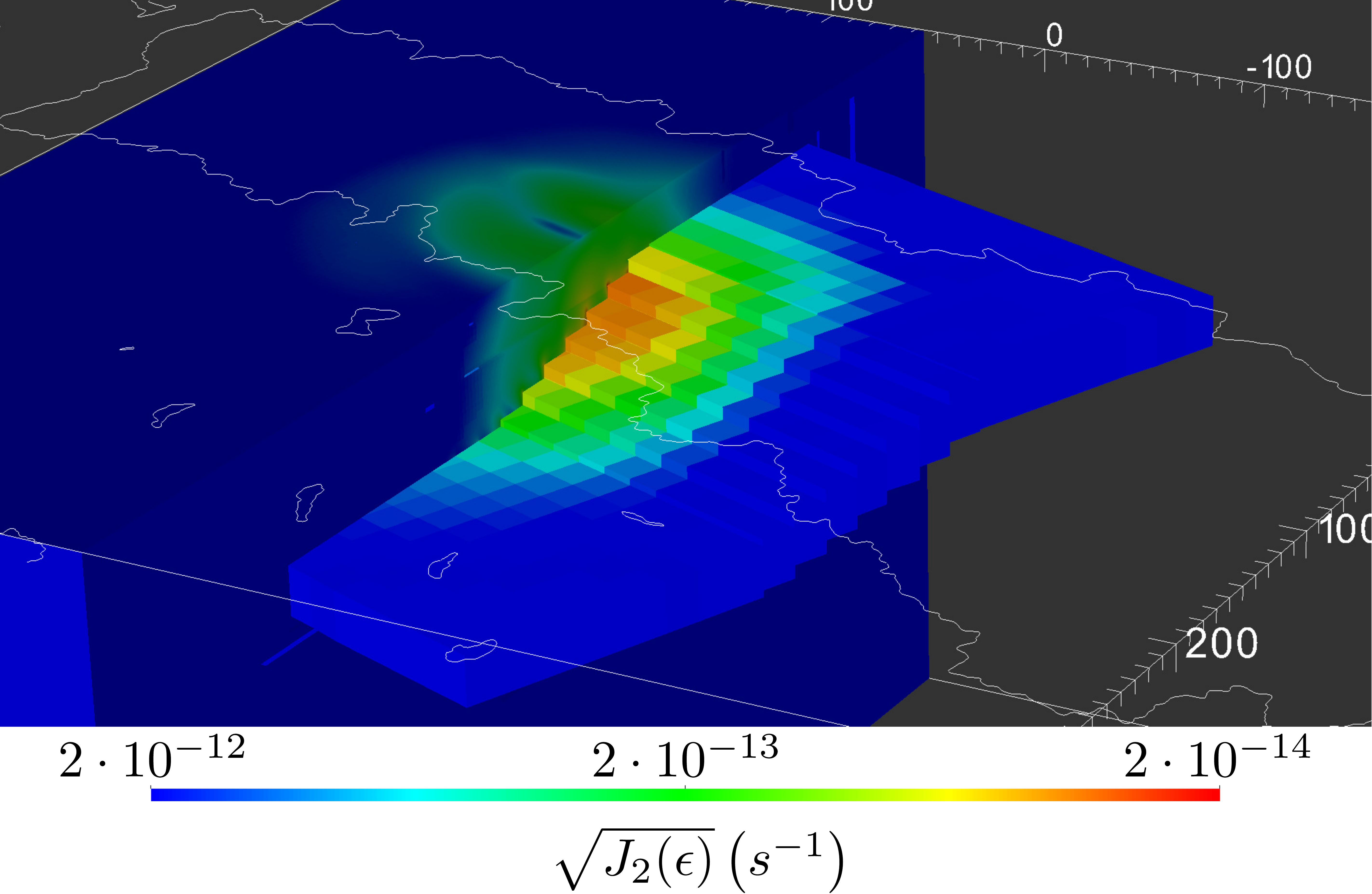}
\par\end{centering}
\caption{\label{fig:Gorkha_strain}A cutout of the applied and instantaneous
induced strain rates for the 2015 Mw 7.8 Gorkha, Nepal earthquake.
We plot the second invariant of the symmetrized strain rate $J_{2}\left(\epsilon\right)=\left(\epsilon_{ij}\epsilon_{ji}-\epsilon_{ii}\epsilon_{jj}/3\right)/2$.
The scale of the model is marked in km, and the political boundaries
of Nepal are superimposed for reference.}
\end{figure}

Using these applied strain rate volumes as input, we use Gamra to
compute the induced strain rate in the bulk (Figure \ref{fig:Gorkha_strain}).
We set the refinement criteria to refine when the error in the induced
velocity is greater than $10^{-11}$ cm/s. This is about 20 times
smaller than the error due to the boundaries being only 300 km away.
Viscoelastic deformation is more distributed than fault slip owing
to the greater depth of the source. The resulting strain rate is more
diffuse, requiring more mesh refinement over a larger volume. This
results in a model that takes longer to solve (26 hours rather than
16 hours for the coseismic slip model).

\section{Conclusions and future work}

We have demonstrated that we can robustly and accurately model arbitrary
distributions of slip on a curvilinear fault and strain in tetrahedral
and cuboidal  volumes. A particular advantage of the adaptive mesh
is that computing the  velocity field due to any single strain volume
is faster than for the overall lower crust model, because the mesh
can remain coarse far away from the source. This makes the approach
well suited for the calculation of elasto-static Green's functions
for localized (e.g., faulting) and distributed (e.g., viscoelastic)
deformation.

While the approach is sufficient for a large class of problems, there
are still some significant limitations. Topography can play a major
influence on the surface displacements when there are large topographic
gradients or if the deformation source is shallow (such as during
volcanic unrest \cite{cayol98,williams+wadge00} or underground explosions).
In addition, the effect of the Earth's curvature can play an important
role for long-wavelength deformation \cite{pollitz96}. Also, coupling
of deformation with the gravity field \cite{pollitz96} is detectable
for particularly large earthquakes \cite{han+06,han+14,vallee+17}.
These areas will be the focus of future work.

\section{Acknowledgements}

This research was supported by the National Research Foundation of
Singapore under the NRF Fellowship scheme (National Research Fellow
Award No. NRF-NRFF2013-04) and by the Earth Observatory of Singapore
and the National Research Foundation and the Singapore Ministry of
Education under the Research Centres of Excellence initiative. This
is EOS publication 184.

\section{Computer Code Availability\label{sec:Computer-Code-Availability}}

The algorithms described in this paper are implemented in Gamra \cite{landry+barbot16},
which is available at \href{https://bitbucket.org/wlandry/gamra}{https://bitbucket.org/wlandry/gamra}.
Gamra is written in C++, uses MPI for parallelism, and depends on
a number of packages: HDF5 \cite{hdf5}, SAMRAI \cite{hornung+02,hornung+06,samrai+online},
FTensor \cite{ftensor+article,ftensor+online}, libokada \cite{libokada},
muparser \cite{muparser-home-page}, and Boost \cite{boost}. Gamra
runs on on everything from laptops to supercomputers. While Gamra
is still under active development, the version associated with this
paper has the Mercurial \cite{mercurial} changeset ID
\begin{lyxcode}
b2d412042fb39cf780ccce431bfb4f476ac74bd7
\end{lyxcode}
\bibliographystyle{abbrv}
\bibliography{reference}

\end{document}